\documentclass[fleqn,usenatbib]{mnras}

\usepackage{newtxtext,newtxmath}

\usepackage[T1]{fontenc}
\usepackage{ae,aecompl}

\usepackage{graphicx}	
\usepackage{amsmath}	
\usepackage{amssymb}	
\usepackage{xcolor}

\title[kHz QPOs and spectral parameters in 4U~1636$-$53]{Relation between the properties of the kilohertz quasi-periodic oscillations and spectral parameters in 4U~1636$-$53.}

\author[E. M. Ribeiro et al.]{
Evandro M. Ribeiro$^{1}$\thanks{E-mail: ribeiro@astro.rug.nl},
Mariano M\'endez$^{1}$,
Guobao Zhang$^{2}$,
and Andrea Sanna$^{3}$
\\
$^{1}$Kapteyn Astronomical Institute, University of Groningen, P.O. BOX 800, 9700 AV Groningen, The Netherlands\\
$^{2}$New York University Abu Dhabi, P.O. Box 129188, Abu Dhabi, United Arab Emirates\\
$^{3}$Dipartimento di Fisica, Universit\'a degli Studi di Cagliari, SP Monserrato-Sestu km 0.7, I-09042, Monserrato, Italy
}

\date{Accepted XXX. Received YYY; in original form ZZZ}

\pubyear{2016}

\begin{document}
\label{firstpage}
\pagerange{\pageref{firstpage}--\pageref{lastpage}}
\maketitle

\begin{abstract}
    We investigate the relation between the parameters of the energy spectrum and the frequency and amplitude of the kilohertz quasi-periodic oscillations (kHz QPOs) in the low-mass X-ray binary 4U~1636$-$53.
    We fit the $3-180$-keV spectrum of this source with a model that includes a thermal Comptonisation component.
    We show that the frequencies of both kHz QPOs follow the same relation as a function of the parameters of this spectral component, except for a systematic frequency shift, whereas the rms fractional amplitude of each QPO follows a different relation with respect to those same parameters.
    This implies that, while the dynamical mechanism that sets the frequencies of the QPO can be the same for both kHz QPOs, the radiative mechanisms that set the amplitudes of the lower and the upper kHz QPO are likely different.
    We discuss the implications of these results to the modelling of the kHz QPOs and the possibility that the lower kHz QPO reflects a resonance between the Comptonising medium and the photons from the accretion disc and/or the neutron star surface.
\end{abstract}

\begin{keywords}
    accretion, accretion discs -- stars: neutron -- X-rays: binaries --
    stars: individual: 4U~1636$-$53
\end{keywords}


\section{Introduction}
\label{sec:intro}

The spectral and timing properties of neutron star low-mass X-ray binaries (NS-LMXB) depend upon the position of the source in the so called colour-colour diagram (CD; e.g., \citealt{Hasinger1989}).
For NS-LMXB the position of the source in the CD can be described by a single parameter, $S_a$ (e.g., \citealt{Kuulkers1994, Mendez1998a}).
The characteristic frequencies observed in the power density spectra (PDS) of these sources are correlated with $S_a$ (e.g., \citealt{Mendez1998a, VanStraaten2000, DiSalvo2001, DiSalvo2003a}).
A typical example is the central frequency of one of the kilohertz quasi-periodic oscillation (kHz QPO), specifically the upper kHz QPO, which is positively correlated with $S_a$ \citep{Mendez1998a, Belloni2005, Belloni2007, Sanna2012}.

\cite{Hasinger1989} proposed that the evolution of the source in the CD is driven by changes in the mass accretion rate in the system and it is often considered a better tracer of accretion rate than X-ray luminosity (e.g., \citealt{Kuulkers1994, Jonker1998, Mendez1998a, Ford2000, VanStraaten2000, Mendez2000}).
As the accretion rate increases the source spectrum softens due to the increase of the relative contribution of the accretion disc and the neutron star surface to the total emission, and the more efficient cooling of the corona.
At the same time, the truncation radius decreases \citep{Done2007} and the temperature at the inner edge of the disc increases.
This scenario is consistent with the correlation between the upper kHz QPO frequency and $S_a$, if this QPO reflects the orbital frequency of matter at the inner edge of the disc \citep{Miller1998,Stella1998}.
The lower kHz QPO appears only within a narrow range of $S_a$ (e.g., \citealt{Zhang2017}), and hence a correlation with $S_a$ is unclear.

4U~1636$-$53 is an LMXB with an orbital period of $3.8$\,h \citep{VanParadijs1990} and the accreting neutron star has a spin rate of 581 Hz \citep{Zhang1997}.
This source shows a cycle of $\sim 40$ days \citep{Shih2005} during which it evolves trough different spectral states covering the full CD. This makes it an exceptional target to explore the relation between its timing and spectral properties.

\cite{Zhang2017} recently studied the relation between the presence of the kHz QPO and the spectral evolution of the source in the NS-LMXB 4U~1636$-$53.
Based on their results, here we study the relation between the properties of the kHz QPO and the spectral properties of this source to gain insight on the mechanism that drives the appearance of the kHz QPOs.
In \S \ref{sec:data} we describe the dataset, models and methods used, in \S \ref{sec:results} we show our results and, finally, we discuss these results and summarise our findings in \S \ref{sec:discussion}

\section{Observations and Data Analysis}
\label{sec:data}

For this work we use all 1576 observations of the NS-LMXB 4U~1636$-$53 taken with the Rossi X-ray Timing Explorer (\textit{RXTE}) Proportional Counter Array (PCA; \citealt{Jahoda2006}) and the High-Energy X-ray Timing Experiment (HEXTE; \citealt{Rothschild1998}).

\subsection{Timing Analysis}
\label{sec:subtiming}

We produced Fourier PDS using the event-mode data with a time resolution of 125 $\mu s$ covering the full PCA energy band.
We set the sample rate to 1/4096\,s, which yields a Nyquist frequency of 2048 Hz, and calculated a PDS every 16\,s; we finally averaged all 16-s PDS of each observation to yield a single PDS per observation.

We ignored frequencies below $200$\,Hz and fitted a Lorentzian to each possible QPO in the average PDS of each observation
using the same criteria as \cite{Sanna2012}, which was to accept all QPOs where the ratio between the Lorentzian normalisation and its negative 1$\sigma$ error was larger than 3, and the coherence, Q, was larger than 2, where Q is the ratio between the central frequency and the FWHM of the Lorentzian.
When only one kHz QPO was detected we identified it as upper or lower kHz QPO based on the branches of the diagram of QPO frequency vs. $S_a$ (see, for example, Figure 1 in \citealt{Sanna2012}).
We detected kHz QPOs in 581 out of 1576 observations; we detected the lower kHz QPO in 403 and the upper in 206 of those observations.

We then combined the data by averaging the PDS of observations where the mean QPO frequency belongs to the frequency ranges shown in \autoref{tab:table1}, for the lower and upper kHz QPO, respectively.
The choice of these frequency ranges was made to allow an approximately equal number of power spectra per frequency range, except for the two lowest frequency ranges of the lower kHz QPO, for which there are only a handful of observations available.
We did not apply any shift-and-add procedure \citep{Mendez1997} to combine the power spectra.

\begin{table*}
	\centering
	\caption{Overview of the selection by QPO frequencies used in this work
    and, for each frequency interval, the average spectral parameters taken
    from \protect\cite{Zhang2017}, and the fractional rms amplitude of the QPO.
    The numbers in between parentheses represent the number of observations used to investigate timing-only properties, before excluding the ones with low values of $\Gamma$ (see text).
    Uncertainties in the rms fractional amplitude represent the 1$\sigma$
    confidence interval obtained from the best fit to the combined power
    spectra. Uncertainties in the other variables represent the 1 $\sigma$
    error of the mean in each averaged interval.}
	\label{tab:table1}
	\begin{tabular}{lccccccc} 
        \hline
        Lower kHz QPO \\
		\hline
         Frequency & \multicolumn{1}{l}{Number of}& \multicolumn{1}{l}{Average} & & & & & \\
		 range (Hz) & observations & frequency (Hz) & $S_a$ & $\Gamma$ & $kT_e$ (keV)& $\tau$ & rms (\%) \\
		$470$--$590$ & $4 $ & $562.1 \pm 8.4$ & $2.013 \pm 0.039$ & $2.13 \pm 0.04$ & $4.26 \pm 0.46$ & $7.6 \pm 0.7$  & $4.27 \pm 0.30$\\
		$590$--$620$ & $4 $ & $608.1 \pm 6.3$ & $2.000 \pm 0.013$ & $2.12 \pm 0.07$ & $4.79 \pm 0.49$ & $7.1 \pm 0.6$  & $4.49 \pm 0.50$\\
		$620$--$670$ & $28\ (30)$ & $648.1 \pm2.3$ & $2.060 \pm 0.005$ & $2.13 \pm 0.05$   & $4.64 \pm 0.30$ & $7.8 \pm 0.6$  & $6.55 \pm 0.15$\\
        $670$--$715$ & $35\ (38)$ & $696.6 \pm2.1$ & $2.093 \pm 0.005$ & $2.05 \pm 0.04$ & $3.86 \pm 0.13$ & $9.1 \pm 0.7$  & $7.56 \pm 0.14$\\
        $715$--$750$ & $36\ (37)$ & $733.0 \pm2.0$ & $2.096 \pm 0.004$ & $2.07 \pm 0.05$ & $3.68 \pm 0.13$ & $9.4 \pm 0.7$  & $7.95 \pm 0.13$\\
        $750$--$790$ & $39\ (42)$ & $766.8 \pm1.7$ & $2.111 \pm 0.004$ & $1.98 \pm 0.04$ & $3.21 \pm 0.08$ & $10.3 \pm 0.4$ & $7.62 \pm 0.10$\\
        $790$--$820$ & $40\ (42)$ & $806.6 \pm1.2$ & $2.118 \pm 0.005$ & $1.90 \pm 0.04$ & $3.09 \pm 0.09$ & $11.2 \pm 0.5$ & $7.74 \pm 0.09$\\
        $820$--$850$ & $45\ (46)$ & $834.5 \pm1.5$ & $2.133 \pm 0.005$ & $1.93 \pm 0.03$ & $3.03 \pm 0.07$ & $10.9 \pm 0.4$ & $7.65 \pm 0.06$\\
        $850$--$880$ & $55\ (59)$ & $864.1 \pm1.2$ & $2.134 \pm 0.004$ & $1.81 \pm 0.03$ & $2.81 \pm 0.04$ & $12.4 \pm 0.4$ & $6.57 \pm 0.05$\\
        $880$--$910$ & $55\ (57)$ & $895.7 \pm1.2$ & $2.142 \pm 0.005$ & $1.77 \pm 0.03$ & $2.78 \pm 0.03$ & $13.1 \pm 0.5$ & $5.48 \pm 0.06$\\
        $910$--$975$ & $44$ & $922.5 \pm1.3$ & $2.149 \pm 0.004$ & $1.81 \pm 0.03$ & $2.85 \pm 0.05$ & $12.4 \pm 0.4$ & $3.98 \pm 0.08$\\
        \hline
        Upper kHz QPO \\
		\hline
        $440$--$540$ & $12 $ & $490.2 \pm9.8$ & $1.336 \pm0.022$ & $1.80 \pm 0.01$ & $9.53  \pm 0.34$ & $6.0 \pm 0.2$  & $13.5 \pm 0.7$\\
        $540$--$650$ & $27 $ & $607.4 \pm4.9$ & $1.509 \pm0.017$ & $1.90 \pm 0.01$ & $12.28 \pm 2.47$ & $5.5 \pm 0.3$  & $11.9 \pm 0.5$\\
        $650$--$750$ & $32$ & $704.9 \pm4.2$ & $1.710 \pm0.009$ & $2.01 \pm 0.02$ & $10.46 \pm 2.01$ & $5.5 \pm 0.2$  & $11.5 \pm 0.3$\\
        $750$--$810$ & $42$ & $783.0 \pm2.4$ & $1.837 \pm0.006$ & $2.14 \pm 0.02$ & $10.13 \pm 1.82$ & $5.3 \pm 0.2$  & $11.4 \pm 0.2$\\
        $810$--$870$ & $28$ & $839.5 \pm3.4$ & $1.918 \pm0.005$ & $2.17 \pm 0.02$ & $7.03 \pm0.79$ & $5.9 \pm 0.3$  & $10.5 \pm 0.2$\\
        $870$--$930$ & $18\ (20)$ & $896.5 \pm4.4$ & $1.989 \pm0.006$ & $2.19 \pm 0.03$ & $5.13 \pm 0.38$ & $6.7 \pm 0.4$ & $9.1 \pm 0.2$\\
        $930$--$1025$ & $17$ & $970.0 \pm6.6$ & $2.054 \pm0.009$ & $2.09 \pm 0.06$ & $4.34 \pm 0.27$ & $8.3 \pm 0.9$ & $6.5 \pm 0.3$\\
        $1025$--$1165$ & $7$ & $1127.5 \pm13.2$ & $2.160 \pm0.015$ & $1.79 \pm 0.06$ & $2.75 \pm 0.03$ & $12.6 \pm 1.0$ & $3.5 \pm 0.3$\\
        $1165$--$1250$ & $18\ (21)$ & $1204.8 \pm4.3$ & $2.194 \pm0.014$ & $1.76 \pm 0.05$ & $2.79 \pm 0.03$ & $13.3 \pm 1.0$ & $2.6 \pm 0.2$\\
        \hline
	\end{tabular}
\end{table*}

For each of the frequency ranges of the lower (upper) QPO we searched the averaged PDS for the corresponding upper (lower) QPO.
This allowed us to recover one of the kHz QPOs from the averaged PDS even if that QPO was not significantly detected in each individual observation.

For each frequency interval of the QPOs we estimated the average background contribution to the PCA lightcurve using the ftool {\sc runpcabackest}, and we used this background count rate, together with that of the source, to calculate the rms amplitude of the QPOs for each selection.

\subsection{Spectral Analysis}
\label{sec:subspec}

For the spectral properties of the source within each frequency interval we used the results from \cite{Zhang2017}.
To get these parameters \cite{Zhang2017} fitted the PCA and HEXTE spectra of 4U~1636$-$53 using the package {\sc Xspec} v12.7 \citep{Arnaud1996}.
To account for interstellar absorption they used the model component {\sc phabs} with cross-sections from \cite{Balucinska-Church1992} and solar abundances from \cite{Anders1989}, fixing the column density to $N_{\rm H} = 3.1 \times 10^{21}$ cm$^{-2}$ \citep{Sanna2013a}.
\cite{Zhang2017} added a multiplicative factor to account for calibration uncertainties between the PCA and HEXTE instruments.

The continuum emission of the source was fitted with a model consisting of a multi-colour disc blackbody ({\sc diskbb} model, \citealt{Mistuda1984, Makishima1986}), a single temperature blackbody ({\sc bbodyrad}) and a Comptonisation model ({\sc nthcomp}, \citealt{Zdziarski1996, Zycki1999}) to describe, respectively, the emission from the disc, the NS surface (or boundary layer), and the Comptonised component.
\cite{Zhang2017} included a Gaussian component with a variable width in the model to account for the presence of an iron line; the energy of the line was fixed to $6.5$ keV.
The temperature of the disc blackbody at the inner disc radius ($k$T$_{\rm dbb}$) was interpolated along the colour-colour diagram using the joint XMM Newton--RXTE spectral fitting results in \cite{Sanna2013a} and set to
0.3, 0.2, 0.4, 0.45, 0.6, 0.75 and 0.8 keV for $S_a$ equal to
 1.1, 1.3, 1.5, 1.7, 1.9, 2.1, and 2.35, respectively. They left the temperature of the {\sc bbodyrad} component free during the fits.

Further details on the spectral models and the fitting results are given in \cite{Zhang2017}. Those authors explored in depth the modelling and interpretation of the X-ray spectra of 4U~1636$-$53 and their relation to the presence of the lower kHz QPO. For this work we use the spectral parameters found by \cite{Zhang2017} to represent the best-fitting model for each observation.

From the Comptonisation model, assuming a spherically symmetric region, the asymptotic power-law photon index, $\Gamma$, can be expressed as:

\begin{equation}
 \label{eq:tau}
 \Gamma  = \left[\frac{9}{4} + \frac{1}{(kT_{e}/m_{e}c^{2})\tau(1 + \tau/3)}\right]^{1/2} - \frac{1}{2},
\end{equation}

\bigskip

where $\tau$ is the optical depth of the region, and $kT_e$ is
the electron temperature \citep{Sunyaev1980}. Both $\Gamma$ and $kT_e$ are obtained from the best fitting model, and are used to calculate the optical depth $\tau$ from \autoref{eq:tau}.

We excluded observations with power-law index $\Gamma < 1.2$ for the averaging of spectral parameters, as that is close to the lower boundary of the allowed parameter values ($\Gamma = 1$) and represents a possible unconstrained fit \citep{Zhang2017}. That left us with a total of 558 observations, containing 385 observations with a lower kHz QPO and 201 with an upper kHz QPO.

\begin{figure}
	\begin{center}
	\includegraphics[width=1.0\columnwidth]{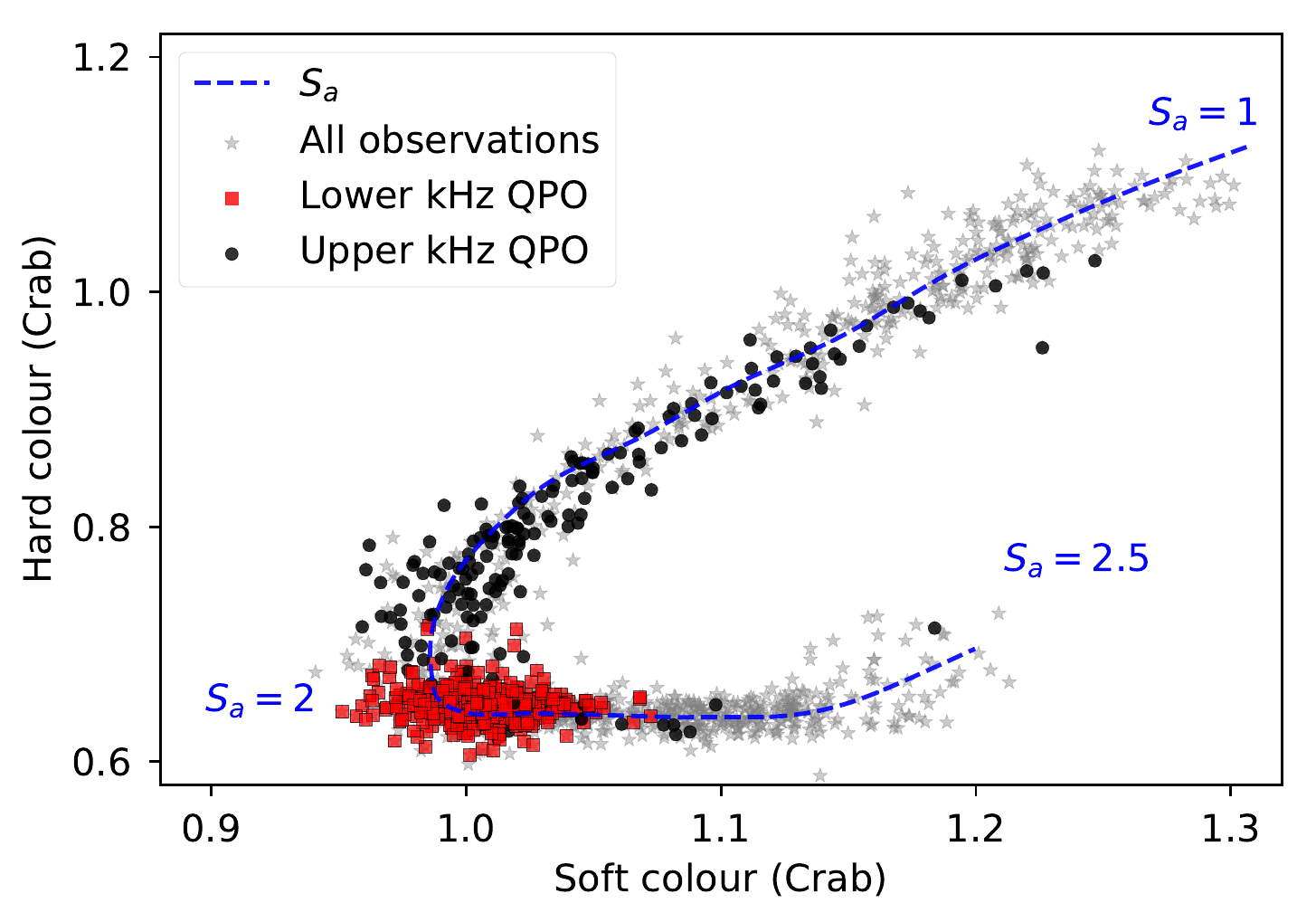}
    \caption{
      Colour-colour diagram of 4U~1636$-$53.
      Each point in this diagram represents a full RXTE/PCA observation. Red and black points represent observations with, respectively, a lower or an upper kHz QPO. (Observations with two simultaneous kHz QPOs are shown in red.) Grey points represent observations without kHz QPOs
      }
    \label{fig:fig1}
	\end{center}
\end{figure}

In order to parameterize the spectral evolution of the source we extracted the PCA lightcurves in 4 energy-bands, following \cite{Zhang2017}, to create a colour-colour diagram (\autoref{fig:fig1}). The hard and soft colours were defined as the count rate ratio between the energy bands $9.7$--$16.0$/$6.0$--$9.7$ keV and $3.5$--$6.0$/$2.0$--$3.5$ keV, respectively, per observation. The parameterization, $S_a$, is defined by the length of the dashed-blue line in \autoref{fig:fig1}, fixing $S_a = 1$ at the top right position of the diagram and $S_a = 2$ at the bottom left position (e.g. \citealt{Mendez1998a}).

\section{Results}
\label{sec:results}

\begin{figure*}
	\begin{center}
	\includegraphics[width=2.0\columnwidth]{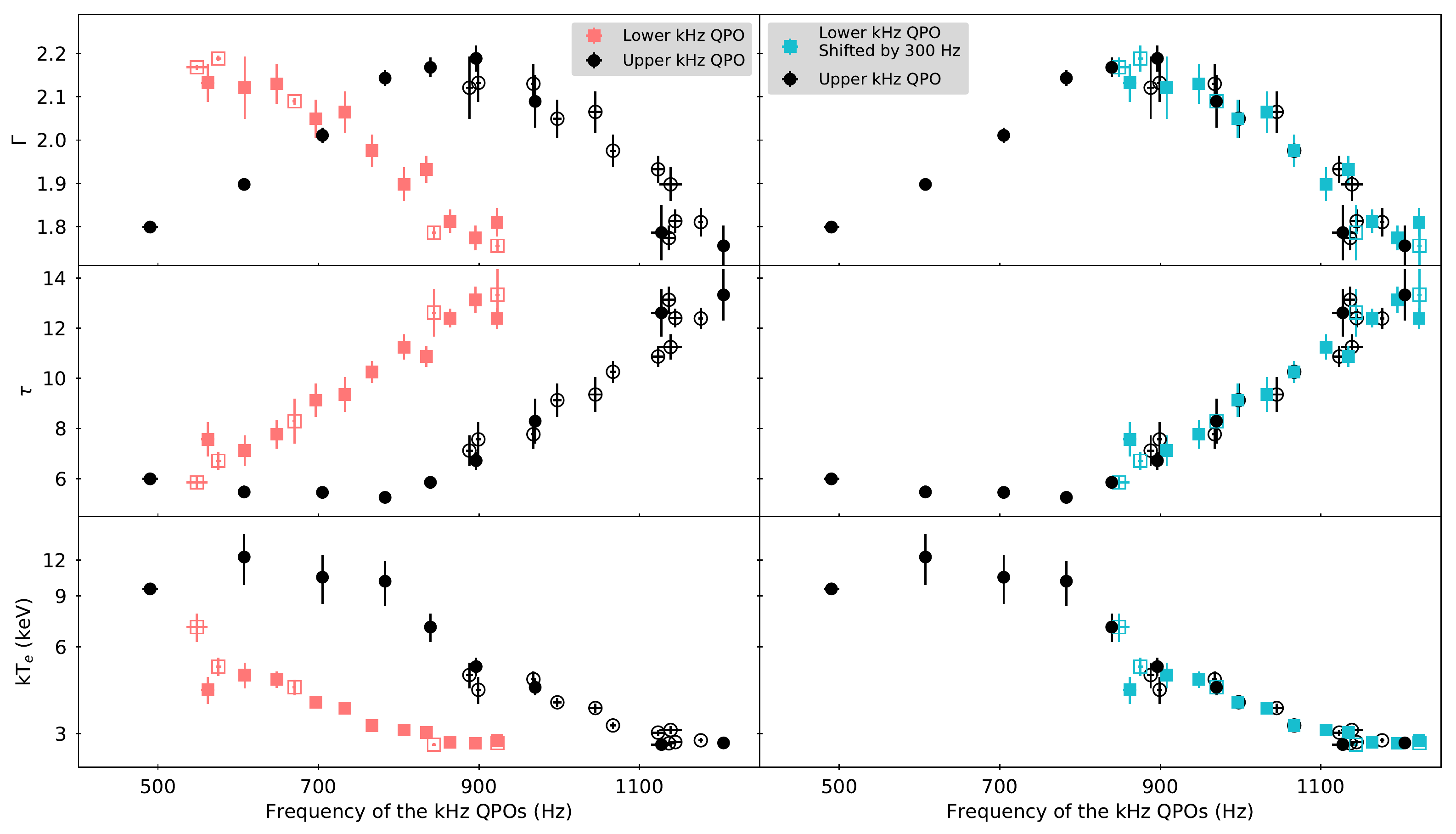}
    \caption{
      Relation between spectral parameters and kHz QPO frequency in 4U~1636$-$53.
      The data are the average in the frequency ranges given in
      \autoref{sec:data}.
      For the plots on the right side the frequency of the lower kHz QPO
      was shifted by 300\,Hz, which is close to the average separation between
      the twin kHz QPO \citep{Jonker2002a}.
      The open black (red or blue) symbols (respectively black and grey in the printed version) correspond to the upper (lower) kHz QPO
      detected after combining the PDSs within the frequency ranges chosen for the
      lower (upper) kHz QPO. Error bars represent the 1$\sigma$ error of the mean in each bin.
      The $y$ axis is shown in \textit{log} scale in the two bottom panels for better visualisation.
      }
    \label{fig:fig2}
	\end{center}
\end{figure*}

 For each of the frequency intervals of the QPO defined in \autoref{tab:table1}, we computed the averaged spectral parameters of the source and we investigated the relation between these averaged parameters and the averaged QPO properties. In \autoref{fig:fig2} we show the relation between averaged spectral parameters and the frequency of the kHz QPOs.
 In all panels red and black (grey and black in the printed version) symbols represent the lower and upper kHz QPO, respectively.

In the upper-left panel of \autoref{fig:fig2} we show the power-law index, $\Gamma$, as a function of QPO frequency.
It is apparent that $\Gamma$ decreases from $\sim 2.2$ to $\sim 1.8$ as the frequency of the lower kHz QPO increases from $\sim 500$ Hz to $\sim 900$ Hz, while for the upper kHz QPO $\Gamma$ first increases from $\sim 1.8$ to $\sim 2.2$ as the frequency of this QPO increases from $\sim 500$ Hz to $\sim 850$ Hz, and then decreases from $\sim 2.2$ to $\sim 1.8$ as the frequency of this QPO increases to approximately $1200$ Hz.

In the middle-left panel of \autoref{fig:fig2} we show the relation between the optical depth, $\tau$, and the frequency of the kHz QPOs.
This figure shows that $\tau$ increases from $6$ to $14$ as the frequency of the lower kHz QPO increases from $\sim 500$ Hz to $\sim 900$ Hz.
As the frequency of the upper kHz QPO increases from $500$ Hz to $850$ Hz $\tau$
remains approximately constant between $5$ and $6$ and then increases to $14$ as the frequency increases from $800$ Hz to $1200$ Hz.

On the bottom-left panel of \autoref{fig:fig2} we show the electron temperature of the Comptonising region, $kT_e$, as a function of QPO frequency.
The temperature remains approximately constant as the frequency of the lower kHz QPO increases from $500$ Hz to $600$ Hz and decreases slowly to $3$ keV as the frequency increases to $900$ Hz.
As the frequency of the upper kHz QPO increases from $400$ Hz to $600$ Hz $kT_e$ remains approximately constant, and then decreases slowly to $3$ keV as the QPO frequency increases to $1200$ Hz

The right-side panels of \autoref{fig:fig2} reproduce the left-side plots
but with the frequency of the lower kHz QPO shifted by 300 Hz, which is close to the average separation between the upper and lower kHz QPO frequencies \citep{Jonker2002a}.
These panels show that the relation between the spectral parameters and the frequency of the upper kHz QPO and the shifted lower kHz QPO can be described by a single relation, as expected due to the almost constant frequency separation between both kHz QPOs. These plots also show that the relations are valid for each QPO regardless of the presence of the other QPO.

\begin{figure*}
	\begin{center}
	\includegraphics[width=1.3\columnwidth]{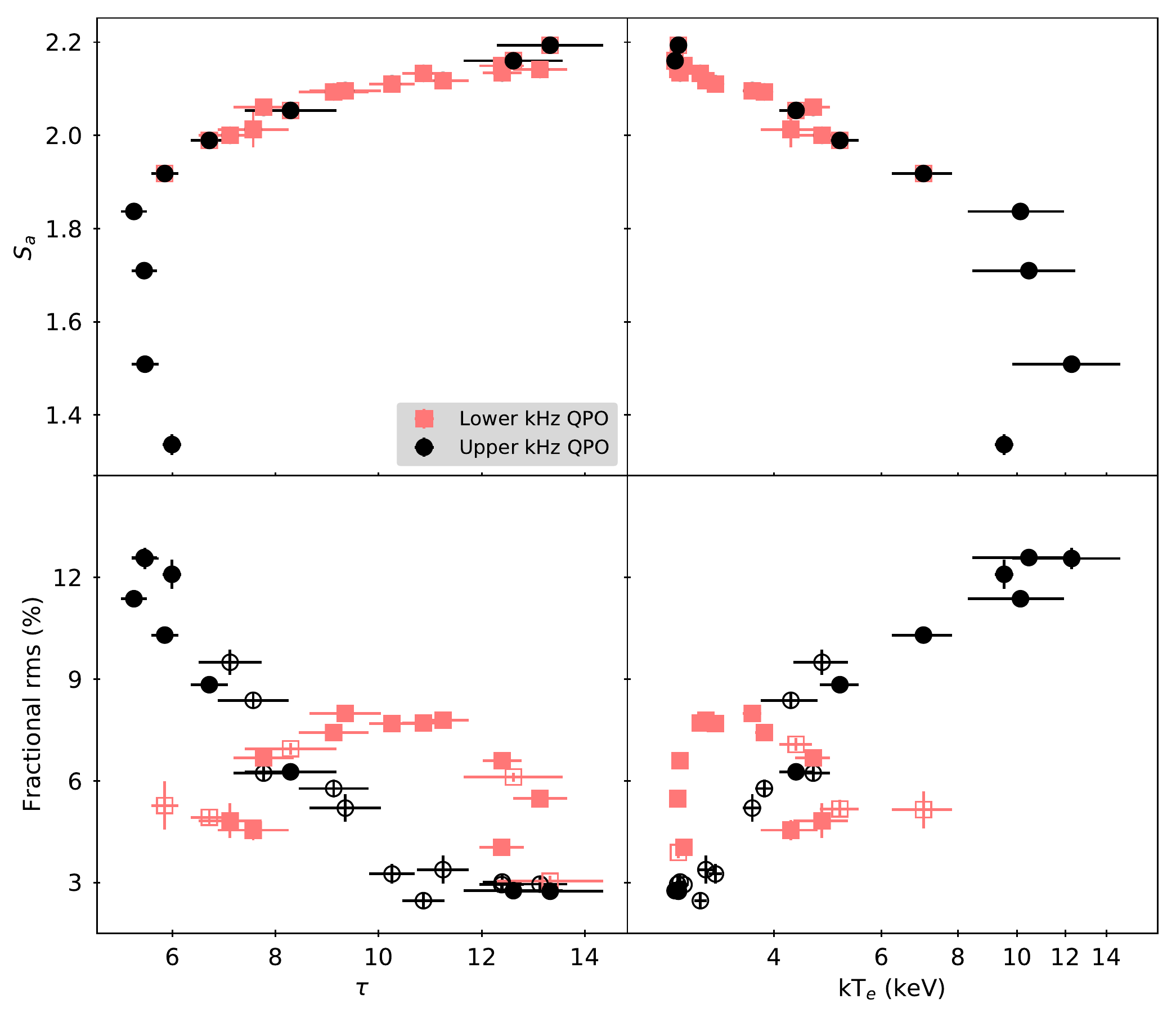}
    \caption{
      Relations of $S_a$ and fractional rms amplitude as a function of spectral parameters in 4U~1636$-$53.
      Symbols are the same as in \autoref{fig:fig2}.
      The $x$ axis of the right side panels are shown in \textit{log} scale for better visualisation.
      Error bars in rms represent the 1$\sigma$  confidence interval obtained
      from the best fit to the combined power spectra. Error bars in the
      other variables represent the 1 $\sigma$ error of the mean in each bin.
    }
    \label{fig:fig3}
	\end{center}
\end{figure*}

In \autoref{fig:fig3} we show the parameter $S_a$\footnotemark~and the fractional rms amplitude of the QPOs as a function of $\tau$ and $kT_{e}$\footnotetext{As explained in \S 1, $S_a$ is an indicator of the state of the source and hence it is usually
displayed in the $x$-axis but here we choose not to follow this convention in order to share both $x$ and $y$ axis for adjacent panels, allowing for a quick comparison between the different relations.}.
In the top-left panel of \autoref{fig:fig3} we show that as $S_a$ increases from $\sim 1.2$ to $\sim 1.8$ the optical depth $\tau$ of the corona remains approximately constant, and then $\tau$ increases steeply as $S_a$ increases from $\sim 1.8$ to $\sim 2.2$.
As noticed by \cite{Zhang2017} the lower kHz QPO is detected only when $\tau$ is larger than $\sim 6$ and increases rapidly.
The relation between $S_a$ and electron temperature $kT_e$ is shown in the top-right panel.
As $S_a$ increases from $\sim 1.2$ to $\sim 1.8$ the temperature is consistent with being constant within errors, and decreases abruptly as $S_a$ increases from $\sim 1.8$ to $\sim 2.2$.
This trend is similar to the one in the top-left panel of this Figure, but one has to keep in mind that the optical depth is obtained from the fitted values of $\Gamma$ and $kT_e$.

The bottom-left panel of \autoref{fig:fig3} shows the fractional rms amplitude of both QPOs as a function of the optical depth $\tau$ of the Comptonising component. When $\tau$ is low ($< 7$) the amplitude of the upper kHz QPO is high ($>10 \%$) and the amplitude decreases to a minimum as $\tau$ increases to $\sim 11$.
The lower kHz QPO displays a very different trend with its rms amplitude increasing from $\sim 4 \%$ to $\sim 8 \%$ as $\tau$ increases from $\sim 5$ to $\sim 11$, and then decreasing as $\tau$ continues increasing to $\sim 14$.

We observe a similar trend in the bottom-right panel showing the relation of the rms amplitude of each QPO with the electron temperature of the Comptonising region.
The amplitude of the upper kHz QPO is low ($\sim 4\%$) when the temperature is below $3$ keV and it increases up to $\sim 14\%$ as the temperature raises to $\sim 15$ keV.
Meanwhile the amplitude of the lower kHz QPO increases from $3 \%$ to $8 \%$ as the temperature increases from $\sim 2.5$ keV to $\sim 3.5$ keV and then decreases to $\sim 5\%$ as the temperature increases up to $\sim 7$ keV.

\begin{figure}
	\begin{center}
	\includegraphics[width=1.0\columnwidth]{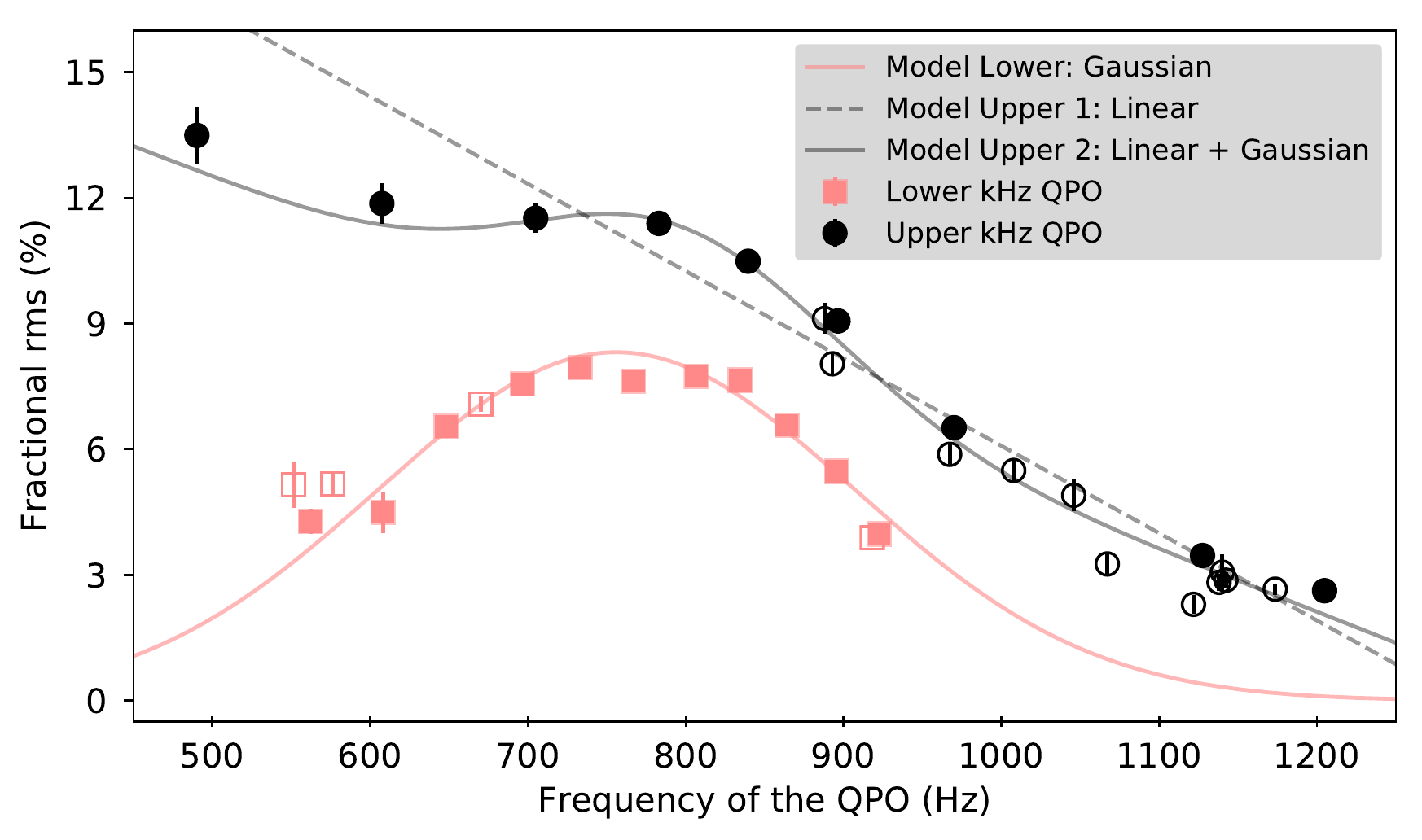}
    \caption{
      The rms amplitude of the upper and lower kHz QPO in 4U 1636$-$53 as a function of the QPO frequency.
      Symbols are the same as in \autoref{fig:fig2}.
      The dashed-black and solid-black lines show, respectively, the best fitting linear function and linear function plus a Gaussian to the rms amplitude of the upper kHz QPO.
      The solid-red (grey in the printed version) line shows the best fitting Gaussian function to the rms amplitude of the lower kHz QPO.
      }
    \label{fig:fig4}
	\end{center}
\end{figure}

\autoref{fig:fig4} shows the rms amplitude vs. frequency of both kHz QPO (see, for example, \citealt{Mendez2000, DiSalvo2003a,Barret2005b, Mendez2006, Altamirano2008a, Boutelier2010a}). The amplitude of the upper QPO decreases from approximately 15\% to 2\% as its frequency increases, with a local maximum ($\sim$ 11\%) around 800 Hz, while the amplitude of the lower QPO increases from 4\% to 8\% as its frequency increases from approximately 500 Hz to 800 Hz and then decreases to 4\% as the frequency increases to 950 Hz.

To quantify the local maximum observed in the plot of the rms amplitude vs. frequency for the upper kHz QPO, we first fitted the data points with a linear function (Model 1) and then with a model consisting of a linear function and a Gaussian (Model 2). We give the best fitting parameters in \autoref{tab:fit} and we plot the data and best-fitting models in \autoref{fig:fig4}.

 Based on Model 1 we simulated $10^7$ datasets with points drawn from a normal distribution with $\sigma$ equal to the $1\sigma$ uncertainties of the rms amplitude at each frequency. We fitted all $10^7$ realisations with Model 2 with the FWHM and central frequency of the Gaussian fixed to the best-fitting values, leaving the amplitude of the Gaussian free.
 From all $10^7$ there was no realisation with an amplitude equal or larger than the one we obtained in Model 2, which implies a probability smaller than $10^{-7}$ that the observed local maximum arises from statistical fluctuations of data that follow a linear relation.

For completeness, and to find the QPO frequency at which the fractional rms amplitude is maximum, we also fitted a Gaussian function to the amplitude of the lower kHz QPO. We show the best-fitting parameters in \autoref{tab:fit}, and we plot the data and the model in \autoref{fig:fig4}.
Our choice of models has no theoretical support and was made simply as a way to quantitatively describe the shape of the rms amplitude of the kHz QPOs as a function of QPO frequency.
The errors of the parameters shown in \autoref{tab:fit} are likely underestimated given that the model is phenomenological and the fits are statistically unacceptable.

\begin{table}
    \centering
    \caption{Best fitting parameters for the rms amplitude of the kHz QPOs in 4U 1636$-$53 as a function of the QPO frequency}
    \label{tab:fit}
    \begin{tabular}{|l l l|}
    \hline
    Upper kHz QPO \\
    & Model 1 : & Model 2:\\
    & Linear & Linear + Gaussian\\
    \hline
    Constant term (\%) & $26.9 \pm 0.3$  & $19 \pm 1$\\
    Slope (\% Hz$^{-1}$)& $-0.0208 \pm 0.0003$  & $-0.0148 \pm 0.0008$\\
    Central frequency (Hz) & -- & $ 799 \pm 11 $ \\
    FWHM (Hz) & -- & $229 \pm 29$ \\
    Amplitude (\%) & -- & $3.2 \pm 0.3$ \\
    $\chi^2 / d.o.f$ & $215.5 / 18$ & $ 65.6 / 15$ \\
    \hline
    Lower kHz QPO \\
    & Model:\\
    & Gaussian\\
    \hline
    Central frequency (Hz)& $755.9 \pm 1.5 $ \\
    FWHM (Hz) & $355 \pm 5 $ \\
    Amplitude (\%) & $8.32 \pm 0.05 $ \\
    $\chi^2 /d.o.f$ & $202.9 / 12$\\
    \hline
  \end{tabular}
\end{table}

\section{Discussion}
\label{sec:discussion}

We show that in the NS-LMXB 4U~1636$-$53 the frequency of the lower and upper kHz QPOs follow the same relation, albeit shifted by $\sim 300$ Hz from each other, as a function of the spectral parameters of the source, but the fractional amplitude of the lower and upper kHz QPOs follow completely different relations.
In the latter case the relation of the rms of one QPO cannot be shifted to obtain the relation of the other.

For frequencies below $\sim 400$ Hz we do not detect
the kHz QPOs, as the PDS of the source starts featuring hectohertz
QPOs and broad-band noise components (see \citealt{Altamirano2008a} for an overview of the X-ray variability in this source).
Therefore it is not possible to tell if Figures \ref{fig:fig2} and \ref{fig:fig3} represent the whole picture.
For example, one might question whether the lower kHz QPO is never present when the temperature $kT_e$ is above $8$ keV, or whether it is simply not detected because its frequency would lie in the noisy region of the PDS.

The frequency of the upper kHz QPO is generally associated with the Keplerian motion of matter at the innermost parts of the accretion disc (e.g. \citealt{Miller1998, Stella1998, Lamb2001, Torok2016}); this idea is consistent with the relation between QPO frequency and $S_a$, which can be understood in terms of changes in the radius at which the QPO is produced due to changes in mass accretion rate (see \S 1).

\autoref{fig:fig2} shows that the dynamical mechanisms that set the frequency of both kHz QPOs are strongly related, or are possibly the same. The frequency of the lower kHz QPO can be shifted to obtain a single relation between the spectral parameters that describe the Comptonising component of the energy spectrum and the kHz QPOs frequencies.
Considering that we observe the upper kHz QPO in different spectral states, and the frequencies of both the upper and lower kHz QPO have the same dependence with spectral parameters, we can argue that, at least dynamically, both kHz QPOs should always be present.
The fact that we do not always detect these QPOs in pairs may be due to differences in the radiative mechanisms that set their amplitudes.

In contrast to the behaviour of the frequencies shown in \autoref{fig:fig2}, \autoref{fig:fig3} shows that the rms fractional amplitude of the upper and lower kHz QPOs have disparate behaviours as a function of spectral parameters. In this case, it is not possible to shift the frequencies or amplitudes of one QPO to obtain the relation shown by the other.
The amplitude of the lower kHz QPO shows a maximum for a specific range of values of the spectral parameters and QPO frequency, whereas the amplitude of the upper kHz QPO decreases more or less monotonically with frequency, with a local maximum (seen as a {\it hump} in \autoref{fig:fig4}) at $\sim 800$ Hz. Therefore the radiative mechanisms that sets the amplitude of, respectively, the lower and the upper kHz QPO are likely different.
It is interesting to notice that the energy and frequency dependence of the time lags of the kHz QPOs in 4U~1636$-$53 also suggest two different radiative mechanisms responsible for each QPO \citep{deAvellar2013}.

As \cite{Zhang2017} pointed out, at $\sim 850$\,Hz the lower kHz is more often detected and has its highest quality factor \citep{Belloni2005, Barret2006a}; this is also the frequency at which the phase lags of the lower kHz QPO are the largest and the coherence between low- and high-energy signals is the highest \citep{deAvellar2013,deAvellar2016}. All these phenomena happening over a relatively narrow range of frequencies, and specific spectral parameters, suggest that a resonance mechanism may drive the amplitude of the lower kHz QPO.

We fitted the rms amplitude of both kHz QPO in order to describe the shape of the rms as a function of the QPO frequency. We find that the local maximum of the amplitude of the upper kHz QPOs and the global maximum of the amplitude of the lower kHz QPO occur respectively at $\sim 800$\,Hz and $\sim 760$\,Hz. Those values were obtained by fitting simple mathematical models to the rms amplitude and they represent the central values of broad features of the rms-frequency relation.

While this may only be a coincidence, it is curious that the peaks of these features are at similar frequencies. (Notice that the spectral properties of the source are different when the lower kHz QPO is at $800$\,Hz than when the upper kHz QPO is at that same frequency.) It could be that a specific combination of spectral parameters (and size, which we do not measure in this work) of the corona triggers (or amplifies) the coupling between disc and corona. One possible scenario is that there is a radiative mechanism that sets the rms amplitude of both QPOs, which peaks at $\sim 800$\,Hz, whereas there is another mechanism that acts only upon the upper kHz QPO and drives the more or less linear decay of its rms with frequency. (Notice that this does not necessarily implies that the QPO frequency drives the rms amplitude, but the rms amplitude could also be determined by the spectral properties of the source.)

The amplitude of both kHz QPOs increases with photon energy (e.g., \citealt{Berger1996, Wijnands1997, Mendez2000, Gilfanov2003}); together with our results this reinforces the notion that even if the oscillations are driven by the dynamics of matter at particular radii in the accretion disc, as for example at the innermost stable circular orbit (e.g, \citealt{Miller1998, Lamb2001, Barret2005a, Barret2005b, Barret2006a, Barret2007a}), the amplitudes of the QPOs are driven by changes in the spectral component that is responsible for the high energy emission in the energy spectrum.
Another possibility is that the oscillations arise from the high-energy component itself (\citealt{Lee2001}; see also \citealt{Kumar2014}), where the Comptonising plasma oscillates at the lower kHz QPO frequency or, as proposed by \cite{Gilfanov2003}, the QPOs reflect oscillations in a Comptonised boundary layer.

Our results support the idea that the properties of the kHz QPOs are the outcome of a complex interplay between the dynamics of matter around the neutron star, and the radiative processes and the interactions between the accretion disc, the neutron star surface, and the Comptonising region.
The fact that frequency and amplitude of the kHz QPO have different behaviours in relation to the spectral parameters should be inherent to any physical model proposed to describe the mechanisms that produce each of the kHz QPOs in 4U 1636$-$53, and possibly other sources.

A key step to understand the nature of the mechanism behind the kHz QPOs
in NS-LMXB is the combined use of spectral and timing techniques to find the link between the timing properties and the physical state of the systems.
We gave here the first steps in this direction by comparing the properties of the kHz QPO and the parameters that describe the Comptonising region on 4U~1636$-$53. Our results should help the development of models that would allow us to describe the physical properties of the Comptonising region, the neutron star itself, and the accretion flow around it, by using the timing properties of the kHz QPO.

\section*{Acknowledgements}

The authors are grateful to the anonymous referee for contributing to the improvement of this paper.
E.M.R acknowledges the support from Conselho Nacional de Desenvolvimento
Cient\'ifico e Tecnol\'ogico (CNPq - Brazil).
This research has made use of data obtained from the High Energy Astrophysics
Science Archive Research Center (HEASARC), provided by NASAs Goddard Space
Flight Center

\bibliographystyle{mnras}
\bibliography{references}

\begin{thebibliography}{}
\makeatletter
\relax
\def\mn@urlcharsother{\let\do\@makeother \do\$\do\&\do\#\do\^\do\_\do\%\do\~}
\def\mn@doi{\begingroup\mn@urlcharsother \@ifnextchar [ {\mn@doi@}
  {\mn@doi@[]}}
\def\mn@doi@[#1]#2{\def\@tempa{#1}\ifx\@tempa\@empty \href
  {http://dx.doi.org/#2} {doi:#2}\else \href {http://dx.doi.org/#2} {#1}\fi
  \endgroup}
\def\mn@eprint#1#2{\mn@eprint@#1:#2::\@nil}
\def\mn@eprint@arXiv#1{\href {http://arxiv.org/abs/#1} {{\tt arXiv:#1}}}
\def\mn@eprint@dblp#1{\href {http://dblp.uni-trier.de/rec/bibtex/#1.xml}
  {dblp:#1}}
\def\mn@eprint@#1:#2:#3:#4\@nil{\def\@tempa {#1}\def\@tempb {#2}\def\@tempc
  {#3}\ifx \@tempc \@empty \let \@tempc \@tempb \let \@tempb \@tempa \fi \ifx
  \@tempb \@empty \def\@tempb {arXiv}\fi \@ifundefined
  {mn@eprint@\@tempb}{\@tempb:\@tempc}{\expandafter \expandafter \csname
  mn@eprint@\@tempb\endcsname \expandafter{\@tempc}}}

\bibitem[\protect\citeauthoryear{Altamirano, van~der Klis, M{\'{e}}ndez,
  Jonker, Klein-Wolt  \& Lewin}{Altamirano et~al.}{2008}]{Altamirano2008a}
Altamirano D.,  van~der Klis M.,  M{\'{e}}ndez M.,  Jonker P.~G.,  Klein-Wolt
  M.,   Lewin W. H.~G.,  2008, \apj, 685, 436

\bibitem[\protect\citeauthoryear{Anders \& Grevesse}{Anders \&
  Grevesse}{1989}]{Anders1989}
Anders E.,  Grevesse N.,  1989, \gca, 53, 197

\bibitem[\protect\citeauthoryear{Arnaud}{Arnaud}{1996}]{Arnaud1996}
Arnaud K.,  1996, Astronomical Data Analysis Software and Systems V, 101, 17

\bibitem[\protect\citeauthoryear{Balucinska-Church \&
  McCammon}{Balucinska-Church \& McCammon}{1992}]{Balucinska-Church1992}
Balucinska-Church M.,  McCammon D.,  1992, \apj, 400, 699

\bibitem[\protect\citeauthoryear{Barret, Olive  \& Miller}{Barret
  et~al.}{2005a}]{Barret2005a}
Barret D.,  Olive J.-F.,   Miller M.~C.,  2005a, Astron. Nachr., 326, 808

\bibitem[\protect\citeauthoryear{Barret, Olive  \& Miller}{Barret
  et~al.}{2005b}]{Barret2005b}
Barret D.,  Olive J.-F.,   Miller M.~C.,  2005b, \mnras, 361, 855

\bibitem[\protect\citeauthoryear{Barret, Olive  \& Miller}{Barret
  et~al.}{2006}]{Barret2006a}
Barret D.,  Olive J.~F.,   Miller M.~C.,  2006, \mnras, 370, 1140

\bibitem[\protect\citeauthoryear{Barret, Olive  \& Miller}{Barret
  et~al.}{2007}]{Barret2007a}
Barret D.,  Olive J.~F.,   Miller M.~C.,  2007, \mnras, 376, 1139

\bibitem[\protect\citeauthoryear{Belloni, M{\'{e}}ndez  \& Homan}{Belloni
  et~al.}{2005}]{Belloni2005}
Belloni T.,  M{\'{e}}ndez M.,   Homan J.,  2005, \aap, 437, 209

\bibitem[\protect\citeauthoryear{Belloni, Homan, Motta, Ratti  \&
  M{\'{e}}ndez}{Belloni et~al.}{2007}]{Belloni2007}
Belloni T.~M.,  Homan J.,  Motta S.,  Ratti E.,   M{\'{e}}ndez M.,  2007,
  \mnras, 379, 247

\bibitem[\protect\citeauthoryear{Berger et~al.,}{Berger
  et~al.}{1996}]{Berger1996}
Berger M.,  et~al., 1996, \apjl, 469, L13

\bibitem[\protect\citeauthoryear{Boutelier, Barret, Lin  \&
  T{\"{o}}r{\"{o}}k}{Boutelier et~al.}{2010}]{Boutelier2010a}
Boutelier M.,  Barret D.,  Lin Y.,   T{\"{o}}r{\"{o}}k G.,  2010, \mnras, 401,
  1290

\bibitem[\protect\citeauthoryear{De~Avellar, M{\'{e}}ndez, Sanna  \&
  Horvath}{De~Avellar et~al.}{2013}]{deAvellar2013}
De~Avellar M. G.~B.,  M{\'{e}}ndez M.,  Sanna A.,   Horvath J.~E.,  2013,
  \mnras, 433, 3453

\bibitem[\protect\citeauthoryear{De~Avellar, M{\'{e}}ndez, Altamirano, Sanna
  \& Zhang}{De~Avellar et~al.}{2016}]{deAvellar2016}
De~Avellar M. G.~B.,  M{\'{e}}ndez M.,  Altamirano D.,  Sanna A.,   Zhang G.,
  2016, \mnras, 461, 79

\bibitem[\protect\citeauthoryear{{Di Salvo}, M{\'{e}}ndez, van~der Klis, Ford
  \& Robba}{{Di Salvo} et~al.}{2001}]{DiSalvo2001}
{Di Salvo} T.,  M{\'{e}}ndez M.,  van~der Klis M.,  Ford E.,   Robba N.~R.,
  2001, \apj, 546, 1107

\bibitem[\protect\citeauthoryear{{Di Salvo}, M{\'{e}}ndez  \& van~der Klis}{{Di
  Salvo} et~al.}{2003}]{DiSalvo2003a}
{Di Salvo} T.,  M{\'{e}}ndez M.,   van~der Klis M.,  2003, \aap, 192, 17

\bibitem[\protect\citeauthoryear{Done, Gierli{\'{n}}ski  \& Kubota}{Done
  et~al.}{2007}]{Done2007}
Done C.,  Gierli{\'{n}}ski M.,   Kubota A.,  2007, \aapr, 15, 1

\bibitem[\protect\citeauthoryear{Ford, van~der Klis, Mendez, Wijnands, Homan,
  Jonker  \& van Paradijs}{Ford et~al.}{2000}]{Ford2000}
Ford E.~C.,  van~der Klis M.,  Mendez M.,  Wijnands R.,  Homan J.,  Jonker
  P.~G.,   van Paradijs J.,  2000, \apj, 537, 368

\bibitem[\protect\citeauthoryear{Gilfanov, Revnivtsev  \& Molkov}{Gilfanov
  et~al.}{2003}]{Gilfanov2003}
Gilfanov M.,  Revnivtsev M.,   Molkov S.,  2003, \aap, 410, 217

\bibitem[\protect\citeauthoryear{Hasinger \& van~der Klis}{Hasinger \& van~der
  Klis}{1989}]{Hasinger1989}
Hasinger G.,  van~der Klis M.,  1989, \aap, 225, 79

\bibitem[\protect\citeauthoryear{Jahoda, Markwardt, Radeva, Rots, Stark, Swank,
  Strohmayer  \& Zhang}{Jahoda et~al.}{2006}]{Jahoda2006}
Jahoda K.,  Markwardt C.~B.,  Radeva Y.,  Rots A.~H.,  Stark M.~J.,  Swank
  J.~H.,  Strohmayer T.~E.,   Zhang W.,  2006, \apjs, 163, 401

\bibitem[\protect\citeauthoryear{Jonker, Wijnands, van~der Klis, Psaltis,
  Kuulkers  \& Lamb}{Jonker et~al.}{1998}]{Jonker1998}
Jonker P.~G.,  Wijnands R.,  van~der Klis M.,  Psaltis D.,  Kuulkers E.,   Lamb
  F.~K.,  1998, \apj, 499, L191

\bibitem[\protect\citeauthoryear{Jonker, M{\'{e}}ndez  \& van~der Klis}{Jonker
  et~al.}{2002}]{Jonker2002a}
Jonker P.~G.,  M{\'{e}}ndez M.,   van~der Klis M.,  2002, \mnras, 336, 1996

\bibitem[\protect\citeauthoryear{Kumar \& Misra}{Kumar \&
  Misra}{2014}]{Kumar2014}
Kumar N.,  Misra R.,  2014, \mnras, 445, 2818

\bibitem[\protect\citeauthoryear{Kuulkers, van~der Klis, Oosterbroek, Asai,
  Dotani, van Paradijs  \& Lewin}{Kuulkers et~al.}{1994}]{Kuulkers1994}
Kuulkers E.,  van~der Klis M.,  Oosterbroek T.,  Asai K.,  Dotani T.,  van
  Paradijs J.,   Lewin W. H.~G.,  1994, \aap, 289, 795

\bibitem[\protect\citeauthoryear{Lamb \& Miller}{Lamb \&
  Miller}{2001}]{Lamb2001}
Lamb F.~K.,  Miller M.~C.,  2001, \apj, 554, 1210

\bibitem[\protect\citeauthoryear{Lee, Misra  \& Taam}{Lee
  et~al.}{2001}]{Lee2001}
Lee H.~C.,  Misra R.,   Taam R.~E.,  2001, \apj, 549, L229

\bibitem[\protect\citeauthoryear{{Makishima}, {Maejima}, {Mitsuda}, {Bradt},
  {Remillard}, {Tuohy}, {Hoshi}  \& {Nakagawa}}{{Makishima}
  et~al.}{1986}]{Makishima1986}
{Makishima} K.,  {Maejima} Y.,  {Mitsuda} K.,  {Bradt} H.~V.,  {Remillard}
  R.~A.,  {Tuohy} I.~R.,  {Hoshi} R.,   {Nakagawa} M.,  1986, \apj, 308, 635

\bibitem[\protect\citeauthoryear{M{\'{e}}ndez}{M{\'{e}}ndez}{2006}]{Mendez2006}
M{\'{e}}ndez M.,  2006, \mnras, 371, 1925

\bibitem[\protect\citeauthoryear{M{\'{e}}ndez et~al.,}{M{\'{e}}ndez
  et~al.}{1998}]{Mendez1997}
M{\'{e}}ndez M.,  et~al., 1998, \apj, 494, L65

\bibitem[\protect\citeauthoryear{M{\'{e}}ndez, van~der Klis, Ford, Wijnands  \&
  van Paradijs}{M{\'{e}}ndez et~al.}{1999}]{Mendez1998a}
M{\'{e}}ndez M.,  van~der Klis M.,  Ford E.~C.,  Wijnands R. a.~D.,   van
  Paradijs J.,  1999, \apj, 511, L49

\bibitem[\protect\citeauthoryear{M{\'{e}}ndez, van~der Klis  \&
  Ford}{M{\'{e}}ndez et~al.}{2001}]{Mendez2000}
M{\'{e}}ndez M.,  van~der Klis M.,   Ford E.~C.,  2001, \apj, 561, 1016

\bibitem[\protect\citeauthoryear{Miller, Lamb  \& Psaltis}{Miller
  et~al.}{1998}]{Miller1998}
Miller M.~C.,  Lamb F.~K.,   Psaltis D.,  1998, \apj, 508, 791

\bibitem[\protect\citeauthoryear{{Mitsuda} et~al.,}{{Mitsuda}
  et~al.}{1984}]{Mistuda1984}
{Mitsuda} K.,  et~al., 1984, \pasj, 36, 741

\bibitem[\protect\citeauthoryear{Rothschild et~al.,}{Rothschild
  et~al.}{1998}]{Rothschild1998}
Rothschild R.~E.,  et~al., 1998, \apj, 496, 538

\bibitem[\protect\citeauthoryear{Sanna, M{\'{e}}ndez, Belloni  \&
  Altamirano}{Sanna et~al.}{2012}]{Sanna2012}
Sanna A.,  M{\'{e}}ndez M.,  Belloni T.~M.,   Altamirano D.,  2012, \mnras,
  424, 2936

\bibitem[\protect\citeauthoryear{Sanna, Hiemstra, M{\'{e}}ndez, Altamirano,
  Belloni  \& Linares}{Sanna et~al.}{2013}]{Sanna2013a}
Sanna A.,  Hiemstra B.,  M{\'{e}}ndez M.,  Altamirano D.,  Belloni T.,
  Linares M.,  2013, \mnras, 432, 1144

\bibitem[\protect\citeauthoryear{Shih, Bird, Charles, Cornelisse  \&
  Tiramani}{Shih et~al.}{2005}]{Shih2005}
Shih I.~C.,  Bird A.~J.,  Charles P.~A.,  Cornelisse R.,   Tiramani D.,  2005,
  \mnras, 361, 602

\bibitem[\protect\citeauthoryear{Stella \& Vietri}{Stella \&
  Vietri}{1998}]{Stella1998}
Stella L.,  Vietri M.,  1998, \apj, 492, L59

\bibitem[\protect\citeauthoryear{Sunyaev \& Titarchuk}{Sunyaev \&
  Titarchuk}{1980}]{Sunyaev1980}
Sunyaev R.~a.,  Titarchuk L.~G.,  1980, \aap, 86, 121

\bibitem[\protect\citeauthoryear{T{\"{o}}r{\"{o}}k, Goluchov{\'{a}},
  Hor{\'{a}}k, {\v{S}}r{\'{a}}mkov{\'{a}}, Urbanec, Pech{\'{a}}{\v{c}}ek  \&
  Bakala}{T{\"{o}}r{\"{o}}k et~al.}{2016}]{Torok2016}
T{\"{o}}r{\"{o}}k G.,  Goluchov{\'{a}} K.,  Hor{\'{a}}k J.,
  {\v{S}}r{\'{a}}mkov{\'{a}} E.,  Urbanec M.,  Pech{\'{a}}{\v{c}}ek T.,
  Bakala P.,  2016, \mnras, 457, L19

\bibitem[\protect\citeauthoryear{Van~Paradijs et~al.,}{Van~Paradijs
  et~al.}{1990}]{VanParadijs1990}
Van~Paradijs J.,  et~al., 1990, \aap, 234, 181

\bibitem[\protect\citeauthoryear{Van~Straaten, Ford, van~der Klis, M{\'{e}}ndez
   \& Kaaret}{Van~Straaten et~al.}{2000}]{VanStraaten2000}
Van~Straaten S.,  Ford E.~C.,  van~der Klis M.,  M{\'{e}}ndez M.,   Kaaret P.,
  2000, \apj, 540, 1049

\bibitem[\protect\citeauthoryear{Wijnands, van~der Klis, van Paradijs, Lewin,
  Lamb, Vaughan  \& Kuulkers}{Wijnands et~al.}{1997}]{Wijnands1997}
Wijnands R. A.~D.,  van~der Klis M.,  van Paradijs J.,  Lewin W. H.~G.,  Lamb
  F.~K.,  Vaughan B.,   Kuulkers E.,  1997, \apj, 479, L141

\bibitem[\protect\citeauthoryear{Zdziarski, Johnson  \& Magdziarz}{Zdziarski
  et~al.}{1996}]{Zdziarski1996}
Zdziarski A.~A.,  Johnson W.~N.,   Magdziarz P.,  1996, \mnras, 283, 193

\bibitem[\protect\citeauthoryear{Zhang, Lapidus, Swank, White  \&
  Titarchuk}{Zhang et~al.}{1997}]{Zhang1997}
Zhang W.,  Lapidus I.,  Swank J.~H.,  White N.~E.,   Titarchuk L.,  1997,
  \iaucirc, 6541

\bibitem[\protect\citeauthoryear{Zhang, M{\'{e}}ndez, Sanna, Ribeiro  \&
  Gelfand}{Zhang et~al.}{2017}]{Zhang2017}
Zhang G.,  M{\'{e}}ndez M.,  Sanna A.,  Ribeiro E.~M.,   Gelfand J.~D.,  2017,
  \mnras, 465, 5003

\bibitem[\protect\citeauthoryear{\.{Z}ycki, Done  \& Smith}{\.{Z}ycki
  et~al.}{1999}]{Zycki1999}
\.{Z}ycki P.~T.,  Done C.,   Smith D.~A.,  1999, \mnras, 309, 561

\makeatother
\end{thebibliography}

\label{lastpage}
\end{document}